\documentclass[preprint]{revtex4}
\usepackage{graphicx}

\begin{document}

\title{$Su(3)$ Algebraic Structure of the Cuprate Superconductors
Model based on the Analogy with Atomic Nuclei}
\author{Shuo Jin$^{1,2,3}$\footnote{Electronic address: jinshuo@buaa.edu.cn},
Bing-Hao Xie$^4$, Hong-Biao Zhang$^5$, Joseph L. Birman$^6$ and
Mo-Lin Ge$^{2,3}$} \affiliation{$^1$Department of Physics, School
of Science, Beihang University, Beijing 100083,
P. R. China\\
$^2$Theoretical Physics Division, Nankai Institute of Mathematics,
Nankai University, Tianjin, 300071, P. R. China\\
$^3$Liuhui Center for Applied Mathematics, Tianjin, 300071, P. R. China\\
$^4$Beijing Information Technology Institute, Beijing, 100101, P.
R. China\\
$^5$Institute of Theoretical Physics, Northeast Normal
University,
Changchun 130024, P.R.China\\
$^6$Department of Physics, The City College of the City University
of New York, New York, 10031, USA}

\begin{abstract}
A cuprate superconductor model based on the analogy with atomic
nuclei was shown by Iachello to have an $su(3)$ structure. The
mean-field approximation Hamiltonian can be written as a linear
function of the generators of $su(3)$ algebra. Using algebraic
method, we derive the eigenvalues of the reduced Hamiltonian
beyond the subalgebras $u(1)\bigotimes u(2)$ and $so(3)$ of
$su(3)$ algebra. In particular, by considering the coherence
between s- and d-wave pairs as perturbation, the effects of
coherent term upon the energy spectrum
are investigated.\\
PACS number(s): 
03.65.Fd,
05.30.Jp
\end{abstract}

\maketitle

\section{introduction} As is well known, the model of
atomic nuclei consisting of $s$ and $d$ nucleon pairs, was
introduced long ago \cite{arima75} and has been very successful in
describing properties of nuclei \cite{arima87}. Recently, it has
been suggested by $M\ddot{u}ller$ \cite{muller1} that both $s$-
and $d$-wave pairing occurs in cuprate superconductors with a
mixture that varies as a function of the distance $r$ from the
surface, which stimulated considerable interests in studying the
novel phenomena of these macroscopic quantum systems
\cite{iachello,muller2,bussmann,furrer,klemm,pan}. A macroscopic
model of cuprate superconductors based on the analogy with atomic
nuclei has been proposed in Ref. \cite{iachello}, and by
developing the OAI method \cite{oai}, Iachello pointed out that
the basic ingredients of the cuprate superconductor model are $s$-
and $d$-wave pairs in a 2-dimensional Fermi system with a surface,
which possesses the properties advocated by $M\ddot{u}ller$
\cite{muller1} on the basis of the analysis of recent experiments.
The analytic solutions of this model based on $u(1)\bigotimes
u(2)$ and $so(3)$ subalgebras structures of the $su(3)$ algebra
have been discussed, which respectively correspond to the case of
ignoring the interaction strength $v_2$ between $s$- and $d$-wave
pairs and choosing $v_2$ as some special value, but the general
solutions going beyond the actual subalgebra-confined framework
have not been investigated. What about the eigenenergy of the
Hamiltonian including the coherence between $s$- and $d$-wave
pairs and the influence of the coherent term on the energy spectra
in this system? This is our task in this paper.

Firstly, we study the cuprate superconductors model where there
appears to be a competition between $s$- and $d$-wave pairing, and
write its Hamiltonian as the bilinear function of the generators
of $su(3)$ algebra. Then we use the mean-field approximate (MFA)
method \cite{solomon} so that the reduced Hamiltonian of this
model can be presented as the linear function of the generators of
$su(3)$ algebra. Adopting the algebraic method \cite{wsj}, we get
the eigenenergies of the reduced Hamiltonian by introducing an
operator of similar transformation. We also discuss the case of
considering the coherence between $s$- and $d$-wave pairs as
perturbation, and analyze the effects of coherent term upon energy
spectrum subsequently.

\section{The model and the reduced Hamiltonian}
By mapping from the fermion Hamiltonian, the boson Hamiltonian of
cuprate superconductors based on the analogy with atomic nuclei
was given by Iachello \cite{iachello}:
\begin{eqnarray}
\label{h1}
H&=&\varepsilon_s(s^+s)+\varepsilon_d(d^+_+d_++d^+_-d_-)+u_0(s^+s^+ss)\nonumber\\
&&+u_2(d^+_+d^+_+d_+d_++d^+_-d^+_-d_-d_-)+u_2'(d^+_+d^+_-d_+d_-)\nonumber\\
&&+v_0(d^+_+d_++d^+_-d_-)(s^+s)+v_2(d^+_+d^+_-ss+s^+s^+d_+d_-),
\end{eqnarray}
where $\varepsilon_s$ and $\varepsilon_d$ denote the single-boson
energies; $u_0,u_2,u_2',v_0$ and $v_2$ are the strengths of the
boson-boson interaction. The creation and annihilation operators
of $s$ and $d_{\pm}$ are denoted by $s^+$, $s$ and $d^+_{\pm}$,
$d_{\pm}$ that satisfy the boson commutation relations:
\begin{eqnarray}
\label{sdr} [s,s^+]=1,\ \ [s,d^+_u]=0,\ \ [s,d_u]=0,\ \
[d_u,d^+_{u'}]=\delta_{uu'},\ \ \ u=\pm.
\end{eqnarray}

An explicit construction of the algebraic generators in terms of
creation and annihilation operators are:
\begin{eqnarray}
&&I_+=d^+_-d_+,\ \ I_-=d^+_+d_-,\nonumber\\
&&U_+=d^+_+s,\ \ \ U_-=s^+d_+,\nonumber\\
&&V_+=s^+d_-,\ \ V_-=d^+_-s,\nonumber\\
&&I_3=\frac{1}{2}(n_{d_-}-n_{d_+}),\nonumber\\
&&I_8=\frac{1}{3}(n_{d_+}+n_{d_-}-2n_{s}),
\end{eqnarray}
where
\begin{eqnarray}
n_s=s^+s,\ \ n_d=n_{d_+}+n_{d_-}=d^+_+d_++d^+_-d_-.
\end{eqnarray}
Based on the relations (\ref{sdr}), one can verify that the set
$\{I_{\pm},I_3,I_8,U_{\pm},V_{\pm}\}$ satisfies the commutation
relations of Lie algebra $su(3)$:
\begin{eqnarray}
\label{re} &&\matrix{ [I_{3},I_{\pm}]=\pm {I_\pm}\ \ \
&[I_{+},I_{-}]=2I_{3}\;\;\;\;\;& [I_{8},I_{\alpha}]=0 \ \ \ \
(\alpha=\pm,3)
\cr\;[I_{3},U_{\pm}]=\mp\frac{1}{2}U_{\pm}
&[I_{8},U_{\pm}]=\pm
U_{\pm}&[U_{+},U_{-}]=-(I_{3}-\frac{3}{2}I_{8})\;\;
\cr
[I_{3},V_{\pm}]=\mp\frac{1}{2}V_{\pm}& [I_{8},V_{\pm}]=\mp V
_{\pm}\;&[V_{+},V_{-}]=-(I_{3}+\frac{3}{2}I_{8})
\cr
[I_{\pm},U_{\pm}]=\pm V_{\mp}\;& [U_{\pm},V_{\pm}]=\pm
I_{\mp}&[I_{\pm},V_{\pm}]=\mp U_{\mp}\;\;\;\;\;\;\;\;\;\;\;}
\nonumber\\
&&\;[I_{\pm},U_{\mp}]=[I_{\pm},V_{\mp}]=[U_{\pm},V_{\mp}]=0.
\end{eqnarray}
Hence the Hamlitonian (\ref{h1}) can be written in terms of the
$su(3)$ generators, i.e.,
\begin{eqnarray}
\label{h2} H
&=&(2u_2-u'_2)I_3^2+(u_0+\frac{u_2}{2}+\frac{u'_2}{4}-v_0)I^2_8
+[\frac{N}{3}(-2u_0+2u_2+u'_2-v_0)-\varepsilon_s+\varepsilon_d+u_0-u_2]I_8\nonumber\\
&&+[\frac{N}{9}(u_0+u'_2+2v_0+2u_2)+\frac{1}{3}(\varepsilon_s+2\varepsilon_d-
u_0-2u_2)]N+v_2(U_+V_-+V_+U_-).
\end{eqnarray}
Here $N=n_s+n_d$ is the total particle number, and $[N,\Gamma]=0\
\ (\Gamma=I_\pm,V_\pm,U_\pm,I_3,I_8)$.

The Hamiltonian (\ref{h2}) can be reduced to the mean field form
by using the standard procedure \cite{solomon} for any bilinear
operators such as $AB$, which can be approximated to the linear
form
\begin{eqnarray}
\label{mfa}
AB \simeq A<B>+<A>B-<A><B>,
\end{eqnarray}
based on the assumption $(A-<A>)(B-<B>)\simeq0$, if $<A>$ and
$<B>$ are suitable expectation values at the ground state.

Setting the expectation values to be:
\begin{eqnarray}
&&<U_+>=\Delta_1,\ \ <U_->=\Delta_1^*,\ \ <V_+>=\Delta_2,\ \
<V_->=\Delta_2^*,\nonumber\\
&&<I_3>=\Delta_3,\ \ <I_8>=\Delta_8,\ \ <N>=N,
\end{eqnarray}
and applying the process (\ref{mfa}) to Hamiltonian (\ref{h2}),
which leads to the mean-field approximation (MFA) of the
Hamiltonian (\ref{h2}):
\begin{eqnarray}
\label{hmf}
&&H_{mf}=H_0+E',\\
\label{h0}
&&H_0=cU_{+}+dU_{-}+eV_{+}+fV_{-}+gI_{3}+hI_{8},
\end{eqnarray}
where the coefficients read
\begin{eqnarray}
\label{c0} &&c=v_2\Delta_2^*,\ \ d=v_2\Delta_2,\ \
e=v_2\Delta_1^*,\ \
f=v_2\Delta_1,\nonumber\\
&&g=2g_0\Delta_3,\ \ h=h_0\Delta_8+h_1N+h_2,\nonumber\\
&&E'=-v_2(\Delta_1\Delta_2^*+\Delta_1^*\Delta_2)-g_0\Delta_3^2-h_0\Delta_8^2+c_1N^2+
(\varepsilon'+h_1\Delta_8+c_2)N-h_1\Delta_8N,\nonumber\\
&&g_0=2u_2-u_2',\ \
h_0=2u_0+u_2+\frac{u_2'}{2}-2v_0,\nonumber\\
&&h_1=\frac{1}{3}(-2u_0+2u_2+u_2'-v_0),\ \
h_2=-\varepsilon_s+\varepsilon_d+u_0-u_2,\nonumber\\
&&c_1=\frac{1}{9}(u_0+u_2'+2v_0+2u_2),\ \
c_2=-\frac{1}{3}(u_0+2u_2),\ \
\varepsilon'=\frac{1}{3}(\varepsilon_s+2\varepsilon_d).
\end{eqnarray}

\section{the energy spectrum}
When the Hamiltonian of a model is expressed as a linear function
of a Lie algebraic generators, there has the algebraic method to
diagonalize and obtain the eigenvalues and eigenstates of the
system. For instance, the algorithm of constructing coherent
states \cite{zwm} and the method of algebraic dynamics
\cite{wsj2}\cite{wsj3}\cite{wsj4}\cite{wsj5} were presented.
Following the standard Lie algebraic theory
\cite{gilmore}\cite{cjq}\cite{humphreys}, if $H_0$ is a linear
function of the generators of a compact semisimple Lie group, it
can be transformed into a linear combination of the Cartan
operators of the corresponding Lie algebra by
\begin{eqnarray}
\label{h'} H_1=WH_0W^{-}.
\end{eqnarray}
Here $W=\prod_{i=1}^Nexp(x_iA_i)$ is an element of the group and
$W^-$ denotes the inverse of $W$, in which {$A_i$} ($i=1,...,N$)
is a basis set in Cartan standard form of the semisimple Lie
algebra, and $x_i$ can be set to zero if the corresponding $A_i$
is a Cartan operator.

In order to give the complete eigenvalues, according to references
\cite{wsj}\cite{zwm}, we choose the similar transition operator as
\begin{eqnarray}
\label{w}
W=exp(x_{21}I_-)exp(x_{23}U_+)exp(x_{31}V_+)exp(x_{32}U_-)exp(x_{13}V_-)exp(x_{12}I_+),
\end{eqnarray}
and let the coefficients of the non-Cartan operator vanish while
substituting Eqs.(\ref{h0})(\ref{w}) into the right-hand side of
Eq.(\ref{h'}), then we get a set of algebraic equations of
$x_{ij}$ after lengthy computation:
\begin{equation} \label{ab3} \left\{
\begin{array}{l}
-gx_{12}+(d-ex_{12})x_{13}=0\\
f+cx_{12}-(h+\frac{1}{2}g)x_{13}-ex^2_{13}=0\\
d-ex_{12}+(h-\frac{1}{2}g+ex_{13}-cx_{32})x_{32}=0,
\end{array}
\right.
\end{equation}
and Hamiltonian after the transformation of $W$ becomes diagonal:
\begin{eqnarray}
\label{whw}
H_1=WH_{0}W^{-}=(g+ex_{13}+cx_{32})I_{3}+(h+\frac{3}{2}ex_{13}-\frac{3}{2}cx_{32})I_{8}.
\end{eqnarray}

One can see that although operator $W$ is not unitary, the similar
transformation (\ref{h'}) guarantees that the eigenvalues of $H_0$
equal those of $H_1$. This is acceptable for we are only concerned
with the eigenvalues. Supposing the common eigenstates of the
Cartan generators $I_3$ and $I_8$ of Lie algebra $su(3)$ are the
Fock states $\mid n_{d_+},n_{d_-},n_s>$, i.e., for the commutative
set $\{I_3,I_8,N\}$ there exist:
\begin{eqnarray}
\label{fock}
&&I_3\mid n_{d_+},n_{d_-},n_s>
=\frac{1}{2}(n_{d_-}-n_{d_+})\mid n_{d_+},n_{d_-},n_s>,\nonumber\\
&&I_8\mid n_{d_+},n_{d_-},n_s>
=\frac{1}{3}(n_{d_+}+n_{d_-}-2n_s)\mid n_{d_+},n_{d_-},n_s>,\nonumber\\
&&N\mid n_{d_+},n_{d_-},n_s> =(n_{d_+}+n_{d_-}+n_s)\mid
n_{d_+},n_{d_-},n_s>.
\end{eqnarray}
From Eqs.(\ref{whw})(\ref{fock}) it follows the eigenvalue of the
MFA Hamiltonian (\ref{hmf}):
\begin{eqnarray}
\label{en} E_{mf}=\frac{1}{2}(g+ex_{13}+cx_{32})(n_{d_-}-n_{d_+})
+\frac{1}{3}(h+\frac{3}{2}ex_{13}-\frac{3}{2}cx_{32})(n_{d_+}+n_{d_-}-2n_s)+E'.
\end{eqnarray}
In the following, we will analyze the solutions of Eq.(\ref{en})
in detail by comparing with Ref. \cite{iachello}.

\subsection{A special case: $v_2=0$}
Considering the case of $v_2=0$, we can easily find that the
Hamiltonian (\ref{h2}) only consists of the $su(3)$ Cartan
operators $\{I_3,I_8\}$ and the total particle number $N$. It is
obvious that the Eq.(\ref{ab3}) only has the zero solution for
$x_{ij}$ and the eigenvalue reads for $v_2=0$:
\begin{eqnarray}
\label{esd}
E^{(A)}(n_s,n_d,l)&=&\varepsilon_sn_s+\varepsilon_dn_d+u_0n_s(n_s-1)+\frac{1}{2}u_2n_d(n_d-2)+\frac{1}{4}u_2'n_d^2\nonumber\\
&&+(2u_2-u_2')l^2+v_0n_dn_s,
\end{eqnarray}
which is the same as ``phase (I)'' corresponding to
$u(1)\bigotimes u(2)$ algebraic structure in Ref.\cite{iachello}.
When the system only has s-wave, the eigenvalue is
$E^{(Aa)}(n_s)=\varepsilon_sn_s+u_0n_s(n_s-1)$, while for p-wave
only, the eigenvalue becomes
$E^{(Ab)}(n_d,l)=\varepsilon_dn_d+\frac{1}{2}u_2n_d(n_d-2)+\frac{1}{4}u_2'n_d^2+(2u_2-u_2')l^2$.
(Note that here $l=-I_3=\frac{1}{2}(n_{d_+}-n_{d_-})$, which is
half of the $l$ defined by Iachello \cite{iachello}.) It can be
naturally understood that Lie algebra $su(3)$ could reduce to its
subalgebra when some parameters are specified.

\subsection{The case of $v_2\neq0$}
To discuss the contribution of coherent term in Hamiltonian
(\ref{h1}), i.e., the nonlinear term in (\ref{h2}), we will
investigate the eigenvalues of the MFA Hamiltonian (\ref{hmf}) by
considering the coherent strength $v_2$ between $s$- and $d$-wave
pairs as a perturbation. For simplicity, we choose $g=h_0=0$,
which means some interaction strengths in Eqs.(\ref{c0}) satisfy
the relations:
\begin{eqnarray}
\label{gh0} u_2'=2u_2,\ \  v_0=u_0+u_2,
\end{eqnarray}
then the coefficients (\ref{c0}) in Hamiltonian (\ref{hmf}) reduce
to
\begin{eqnarray}
\label{jh} &&c=v_2\Delta_2^*,\ \ d=v_2\Delta_2,\ \
e=v_2\Delta_1^*,\ \
f=v_2\Delta_1,\nonumber\\
&&g=0,\nonumber\ \ h=h_1N+h_2,\nonumber\\
&&E'=-v_2(\Delta_1\Delta_2^*+\Delta_1^*\Delta_2)+c_1N^2+(\varepsilon'+h_1\Delta_8+c_2)N-h_1\Delta_8N,\nonumber\\
&&h_1=u_2-u_0,\ \
h_2=-\varepsilon_s+\varepsilon_d+u_0-u_2,\nonumber\\
&&c_1=\frac{1}{3}(u_0+2u_2),\ \ c_2=-\frac{1}{3}(u_0+2u_2),\ \
\varepsilon'=\frac{1}{3}(\varepsilon_s+2\varepsilon_d),
\end{eqnarray}
and the solutions of Eqs.(\ref{ab3}) read:
\begin{eqnarray}
\label{x6} 1.&&\ \ x_{13}=0,\ \ x_{12}=-\frac{f}{c},\ \
x_{32}=\frac{1}{2c}(h-\lambda),\nonumber\\
2.&&\ \ x_{13}=0,\ \ x_{12}=-\frac{f}{c},\ \
x_{32}=\frac{1}{2c}(h+\lambda),\nonumber\\
3.&&\ \ x_{13}=-\frac{1}{2e}(h+\lambda),\ \ x_{12}=\frac{d}{e},\ \
x_{32}=0,\nonumber\\
4.&&\ \ x_{13}=-\frac{1}{2e}(h-\lambda),\ \ x_{12}=\frac{d}{e},\ \
x_{32}=0,\nonumber\\
5.&&\ \ x_{13}=\frac{1}{2e}(h+\lambda),\ \ x_{12}=\frac{d}{e},\ \
x_{32}=\frac{1}{2e}(h-\lambda),\nonumber\\
6.&&\ \ x_{13}=\frac{1}{2e}(h-\lambda),\ \ x_{12}=\frac{d}{e},\ \
x_{32}=\frac{1}{2e}(h+\lambda),
\end{eqnarray}
where
\begin{eqnarray}
\label{v}
\lambda=\sqrt{h^2+4v_2^2(\Delta_1\Delta_1^*+\Delta_2\Delta_2^*)}.
\end{eqnarray}
Substituting Eq.(\ref{x6}) into Eq.(\ref{en}), we obtain the
eigenvalues of reduced Hamiltonian (\ref{hmf}) as:
\begin{eqnarray}
\label{e6}
&&E_1=-(\frac{h-3\lambda}{6})n_{d_{+}}+\frac{h}{3}n_{d_-}-(\frac{h+3\lambda}{6})n_{s}+E',\nonumber\\
&&E_2=-(\frac{h+3\lambda}{6})n_{d_{+}}+\frac{h}{3}n_{d_-}-(\frac{h-3\lambda}{6})n_{s}+E',\nonumber\\
&&E_3=\frac{h}{3}n_{d_{+}}-(\frac{h+3\lambda}{6})n_{d_-}-(\frac{h-3\lambda}{6})n_{s}+E',\nonumber\\
&&E_4=\frac{h}{3}n_{d_{+}}-(\frac{h-3\lambda}{6})n_{d_-}-(\frac{h+3\lambda}{6})n_{s}+E',\nonumber\\
&&E_5=-(\frac{h-3\lambda}{6})n_{d_{+}}-(\frac{h-3\lambda}{6})n_{d_-}+\frac{h}{3}n_{s}+E',\nonumber\\
&&E_6=-(\frac{h+3\lambda}{6})n_{d_{+}}-(\frac{h+3\lambda}{6})n_{d_-}+\frac{h}{3}n_{s}+E'.
\end{eqnarray}

If coherent coefficient $v_2$ is much smaller than others in the
Hamiltonian (\ref{hmf}), we regard the nonlinear term including
$v_2$ as a perturbation one. For $v_2\ll h$ by expanding $\lambda$
to the second-order, we get
\begin{eqnarray}
\label{v1} \lambda\simeq\pm
h[1+\frac{2v_2^2}{h^2}(\Delta_1\Delta_1^*+\Delta_2\Delta_2^*)].
\end{eqnarray}
Substituting (\ref{v1}) into (\ref{e6}) and supposing
$n_{d_{+}}=n_{d_{-}}$ in the following, we obtain
\begin{enumerate}

\item{$
E_1=-(\frac{h-3\lambda}{6})n_{d_{+}}+\frac{h}{3}n_{d_-}-(\frac{h+3\lambda}{6})n_{s}+E'\
\ (x_{13}=0,\ \ x_{12}=-\frac{f}{c},\ \
x_{32}=\frac{1}{2c}(h-\lambda)) $}
\begin{enumerate}
\item{For $h>0$,}
\begin{eqnarray}
\label{e1a}
E_{1a}&=&E^{(I)}+\alpha v_2^2(n_{d_+}-n_s)-\beta v_2,\nonumber\\
&=&E^{(I)}+\alpha v_2^2(\frac{n_{d}}{2}-n_s)-\beta v_2;
\end{eqnarray}

\item{when $h<0$}
\begin{eqnarray}
\label{e1b}
E_{1b}&=&E^{(II)}+\frac{h}{3}(-2n_{d_+}+n_{d_-}+n_s)-\alpha v_2^2n_{d_+}+\alpha v_2^2n_s-\beta v_2,\nonumber\\
&=&E^{(II)}-(\frac{h}{3}+\alpha v_2^2)(\frac{n_d}{2}-n_s)-\beta
v_2,
\end{eqnarray}
where
\begin{eqnarray}
\label{e11}
&&E^{(I)}=\varepsilon_sn_s+\varepsilon_dn_d+u_0n_s(n_s-1)+u_2n_d(n_d-1)+(u_0+u_2)n_sn_d,\\
\label{e12}
&&\alpha=\frac{\Delta_1\Delta_1^*+\Delta_2\Delta_2^*}{(u_2-u_0)N-\varepsilon_s+\varepsilon_d+u_0-u_2},\
\ \  \beta=\Delta_1\Delta_2^*+\Delta_1^*\Delta_2,
\end{eqnarray}
and
\begin{eqnarray}
\label{e21}
E^{(II)}=\frac{1}{3}(u_0+2u_2)N^2+\frac{1}{3}(\varepsilon_s+2\varepsilon_d-u_0-2u_2)N.
\end{eqnarray}
According to Eqs.(\ref{jh})(\ref{e12}), $h>0$, or $\alpha>0$,
corresponds to the case of $u_2>u_0$ and
$\varepsilon_d>\varepsilon_s$, that is, the energy of $s$-wave is
lower than that of $d$-wave, and the self-interaction strength of
$d$-wave is larger than that of $s$-wave. On the contrary, $h<0$,
or $\alpha<0$, is $u_2<u_0$ and $\varepsilon_d<\varepsilon_s$, the
energy of $d$-wave is lower than that of $s$-wave, and the
self-interaction strength of $s$-wave is larger than that of
$d$-wave. $\beta$ denotes the part of the contribution of
coherence between $s$-wave and $d$-wave and is positive.

From Eqs.(\ref{e1a})(\ref{e1b}), one can draw the conclusion that
when $v_2=0$, the eigenenergy $E_{1a}$ reduces to $E^{(I)}$, the
eigenenergy of $s$-wave and $d$-wave mixed without the coherent
interaction. Note that when $n_{d_{+}}=n_{d_-}$ and Eq.(\ref{gh0})
holds, the eigenenergy $E^{(A)}$ (\ref{esd}) reduces to $E^{(I)}$
(\ref{e11}). Comparing with Ref.\cite{iachello}, $E_{1b}$ gives
another eigenvalue of the system, and when $v_2=0$ $E_{1b}$
reduces to $E^{(II)}$, which is another eigenenergy of $s$-wave
and $d$-wave mixed without the coherent interaction. Based on
Eqs.(\ref{e11})(\ref{e21}), we find that $E^{(I)}>E^{(II)}$ when
$\alpha>0$ and $2n_s<n_d$ or $\alpha<0$ and $2n_s>n_d$;
$E^{(I)}<E^{(II)}$ holds for $\alpha>0$ and $2n_s>n_d$ or
$\alpha<0$ and $2n_s<n_d$, i.e., there exists the case that
$E^{(I)}$ or $E^{(II)}$ is the ground state for different cases.
The last two terms in Eq.(\ref{e1a}) and Eq.(\ref{e1b}) are the
contributions of the coherent term indicated by $v_2$ in
Hamiltonian (\ref{hmf}) corresponding to eigenenergies $E_{1a}$
and $E_{1b}$ respectively, the first term including $v_2$ causes
split between $n_{d}$ and $n_s$ energy levels, and the last term
leads to the energy have a total shift, whose direction is
determined by the sign of $v_2$. Split between $n_{d}$ and $n_{s}$
and the total shift make the energy spectrum different from the
case of $v_2=0$, especially the change of the energy gap between
$s$-wave and $d$-wave.

For $h>0$ and $v_2>0$, the energy eigenvalue $E_{1a}$ has the
smaller value when the particle number $n_s$ is more than the
particle number $n_d$, and the coherent strength between $s$-wave
and $d$-wave is large as possible. For the case of $h<0$ and
$v_2>0$, the larger $n_s$ takes and the larger coherence between
$s$-wave and $d$-wave has, the more steady the system is.
\end{enumerate}

\item{$
E_2=-(\frac{h+3\lambda}{6})n_{d_{+}}+\frac{h}{3}n_{d_-}-(\frac{h-3\lambda}{6})n_{s}+E'
\ \ (x_{13}=0,\ \ x_{12}=-\frac{f}{c},\ \
x_{32}=\frac{1}{2c}(h+\lambda))$}
\begin{enumerate}
\item{For $h>0$,}
\begin{eqnarray}
E_{2a}&=&E^{(II)}+\frac{h}{3}(-2n_{d_+}+n_{d_-}+n_s)-\alpha v_2^2n_{d_+}+\alpha v_2^2n_s-\beta v_2,\nonumber\\
&=&E^{(II)}-(\frac{h}{3}+\alpha v_2^2)(\frac{n_d}{2}-n_s)-\beta
v_2.
\end{eqnarray}
\item{When $h<0$,}
\begin{eqnarray}
E_{2b}&=&E^{(I)}+\alpha v_2^2(n_{d_+}-n_s)-\beta v_2,\nonumber\\
&=&E^{(I)}+\alpha v_2^2(\frac{n_{d}}{2}-n_s)-\beta v_2.
\end{eqnarray}
Similar to the analysis of case 1, we can find that for $h>0$ and
$v_2>0$, the system becomes more steady when both $n_d$ and the
coherent strength are larger. And when $h<0$, the more $n_d$ and
the coherent strength takes, the lower the energy eigenvalue has
for $v_2>0$.
\end{enumerate}

\item{$
E_3=\frac{h}{3}n_{d_{+}}-(\frac{h+3\lambda}{6})n_{d_-}-(\frac{h-3\lambda}{6})n_{s}+E'
\ \ (x_{13}=-\frac{1}{2e}(h+\lambda),\ \ x_{12}=\frac{d}{e},\ \
x_{32}=0)$}
\begin{enumerate}
\item{For $h>0$,}
\begin{eqnarray}
E_{3a}&=&E^{(II)}+\frac{h}{3}(n_{d_+}-2n_{d_-}+n_s)-\alpha v_2^2n_{d_-}+\alpha v_2^2n_s-\beta v_2,\nonumber\\
&=&E^{(II)}-(\frac{h}{3}+\alpha v_2^2)(-\frac{n_d}{2}-n_s)-\beta
v_2.
\end{eqnarray}
\item{When $h<0$,}
\begin{eqnarray}
E_{3b}&=&E^{(I)}+\alpha v_2^2(n_{d_-}-n_s)-\beta v_2,\nonumber\\
&=&E^{(I)}+\alpha v_2^2(\frac{n_{d}}{2}-n_s)-\beta v_2.
\end{eqnarray}
For $h>0$ and $v_2>0$, the system becomes more steady when both
$n_d$ and the coherent strength are larger. And when $h<0$, the
more $n_d$ and the coherent strength takes, the lower the energy
eigenvalue has for $v_2>0$.
\end{enumerate}

\item{$
E_4=\frac{h}{3}n_{d_{+}}-(\frac{h-3\lambda}{6})n_{d_-}-(\frac{h+3\lambda}{6})n_{s}+E'
\ \ (x_{13}=-\frac{1}{2e}(h-\lambda),\ \ x_{12}=\frac{d}{e},\ \
x_{32}=0)$}
\begin{enumerate}
\item{For $h>0$,}
\begin{eqnarray}
E_{4a}&=&E^{(I)}+\alpha v_2^2(n_{d_-}-n_s)-\beta v_2,\nonumber\\
&=&E^{(I)}+\alpha v_2^2(\frac{n_{d}}{2}-n_s)-\beta v_2.
\end{eqnarray}
\item{When $h<0$,}
\begin{eqnarray}
E_{4b}&=&E^{(II)}+\frac{h}{3}(n_{d_+}-2n_{d_-}+n_s)-\alpha v_2^2n_{d_-}+\alpha v_2^2n_s-\beta v_2,\nonumber\\
&=&E^{(II)}-(\frac{h}{3}+\alpha v_2^2)(-\frac{n_d}{2}-n_s)-\beta
v_2.
\end{eqnarray}
When $h>0$, the more $n_s$ and the coherent strength takes, the
lower the energy eigenvalue has for $v_2>0$. For $h<0$ and
$v_2>0$, the system becomes more steady when both $n_s$ and the
coherent strength are larger.
\end{enumerate}

\item{$
E_5=-(\frac{h-3\lambda}{6})n_{d_{+}}-(\frac{h-3\lambda}{6})n_{d_-}+\frac{h}{3}n_{s}+E'
\ \ (x_{13}=\frac{1}{2e}(h+\lambda),\ \ x_{12}=\frac{d}{e},\ \
x_{32}=\frac{1}{2e}(h-\lambda)$}
\begin{enumerate}
\item{For $h>0$,}
\begin{eqnarray}
E_{5a}&=&E^{(II)}+\frac{h}{3}N+\alpha v_2^2(n_{d_+}+n_{d_-})-\beta v_2,\nonumber\\
&=&E^{(II)}+\alpha v_2^2n_{d}-\beta v_2.
\end{eqnarray}
\item{When $h<0$,}
\begin{eqnarray}
E_{5b}&=&E^{(II)}+\frac{h}{3}(-2n_{d_+}-2n_{d_-}+n_s)-\alpha v_2^2(n_{d_+}+n_{d_-})-\beta v_2,\nonumber\\
&=&E^{(II)}-(\frac{2h}{3}+\alpha v_2^2)n_d+\frac{h}{3}n_s-\beta
v_2.
\end{eqnarray}
Both for $h>0$ and $h<0$, the eigenenergies give new values, and
reduce to $E^{(II)}$ when $v_2=0$. When $h>0$ and $v_2>0$, the
less $n_d$ has and the larger the coherent strength takes, the
more steady the system is. For $h<0$ and $v_2>0$, the more $n_s$
has and the larger the coherent strength takes, the more steady
the system is.
\end{enumerate}

\item{$
E_6=-(\frac{h+3\lambda}{6})n_{d_{+}}-(\frac{h+3\lambda}{6})n_{d_-}+\frac{h}{3}n_{s}+E'
\ \ (x_{13}=\frac{1}{2e}(h-\lambda),\ \ x_{12}=\frac{d}{e},\ \
x_{32}=\frac{1}{2e}(h+\lambda))$}
\begin{enumerate}
\item{For $h>0$,}
\begin{eqnarray}
E_{6a}&=&E^{(II)}+\frac{h}{3}(-2n_{d_+}-2n_{d_-}+n_s)-\alpha v_2^2(n_{d_+}+n_{d_-})-\beta v_2,\nonumber\\
&=&E^{(II)}-(\frac{2h}{3}+\alpha v_2^2)n_d+\frac{h}{3}n_s-\beta
v_2.
\end{eqnarray}
\item{When $h<0$,}
\begin{eqnarray}
E_{6b}&=&E^{(II)}+\frac{h}{3}N+\alpha v_2^2(n_{d_+}+n_{d_-})-\beta v_2,\nonumber\\
&=&E^{(II)}+\alpha v_2^2n_{d}-\beta v_2.
\end{eqnarray}
Both for $h>0$ and $h<0$, the eigenenergies give new values, and
reduce to $E^{(II)}$. Either when $h>0$ and $v_2>0$ or for the
case of $h<0$ and $v_2>0$, the more $n_d$ has and the larger the
coherent strength takes, the more steady the system is.
\end{enumerate}
\end{enumerate}

\section{conclusion}
We have considered a cuprate superconductor model with $s$- and
$d$-wave pairs which exhibits a dynamical $su(3)$ symmetry. In
order to investigate the solutions of this quantum system, we have
used the mean-field approximation and got the reduced Hamiltonian
in a form with linear function of the $su(3)$ algebra generators.
By introducing the similar transformation operator $W$ and
eliminating the non-Cartan generators after the similar
transformation through parameter $x_{ij}$, we have obtained the
eigenenergies of the reduced Hamiltonian beyond the
subalgebra-confined framework. It should be noted that the order
of the operators in $W$ can be chosen arbitrarily, but the
coefficients $x_{ij}$ are strongly dependent on the order, that
is, the operator $W$ in this algebraic method is not exclusive.
Although any specified order has a solution, a properly chosen
order can simplify the procedure to get the $x_{ij}$. The
advantage of this algebraic method is that the transformation
operator $W$ ensures the eigenvalue of Hamiltonian $H_1$ as the
same as that of Hamiltonian $H$. In particular, we have analyzed
the effects of the coherent term, when we consider it as a
perturbative one. We get another new eigenvalue $E^{(II)}$ for the
case of $s$-wave and $d$-wave mixed without the coherent
interaction. Furthermore, the split and the whole shift of the
energy spectra induced by the coherent term could be understood
clearly. Much work (such as the numerical calculations) remains to
be done in the field of searching and seeking more general
solutions going beyond the present works, which can help us to
understand more and deeper about the model of cuprate
superconductors based on the analogy with atomic nuclei.

\begin{acknowledgments} This work is in part supported by
NSF of China under Grant No. 10447103, No. 10405006, and the
Foundation of Beihang University.
\end{acknowledgments}

\end{document}